\documentclass[preprint,showpacs,preprintnumbers,amsmath,amssymb]{revtex4}
\usepackage{graphicx}
\usepackage{dcolumn}
\usepackage{bm}
\begin{document}

\newcommand{\re}{\mathop{\mathrm{Re}}}
\newcommand{\im}{\mathop{\mathrm{Im}}}
\newcommand{\D}{\mathop{\mathrm{d}}}
\newcommand{\I}{\mathop{\mathrm{i}}}
\newcommand{\E}{\mathop{\mathrm{e}}}

\title{
SASE FEL with energy-chirped electron beam and its application for
generation of attosecond pulses
}\thanks{\normalsize Preprint DESY 06-051, April 2006 \hfill
submitted to Phys. Rev. ST AB}

\author{E.L.~Saldin, E.A.~Schneidmiller,  and M.V.~Yurkov}

\affiliation{Deutsches Elektronen-Synchrotron (DESY),
Hamburg, Germany}


\begin{abstract}

Influence of a linear energy chirp in the electron beam on a SASE FEL operation
is studied analytically and numerically using 1-D model. Explicit expressions
for Green's functions and for output power of a SASE FEL are obtained for
high-gain linear regime in the limits of small and large energy chirp
parameter. Saturation length and power versus energy chirp parameter are
calculated numerically. It is shown that the effect of linear energy chirp
on FEL gain is equivalent to the linear undulator tapering (or linear energy
variation along the undulator). A consequence of this fact is a possibility to
perfectly compensate FEL gain degradation, caused by the energy chirp, by means
of the undulator tapering independently of the value of the energy chirp parameter.
An application of this effect for generation of attosecond pulses from a hard X-ray
FEL is proposed. Strong energy modulation within a short slice of an electron bunch is
produced by few-cycle optical laser pulse in a short undulator, placed in front of the main undulator.
Gain degradation within this slice is
compensated by an appropriate undulator taper while the rest of the bunch suffers from this taper
and does not lase. Three-dimensional simulations predict that
short (200 attoseconds) high-power (up to 100 GW) pulses can be produced in Angstroem wavelength
range with a high degree of contrast. A possibility to reduce pulse duration to sub-100
attosecond scale is discussed.

\end{abstract}


\maketitle

\section{Introduction}

Start-to-end simulations \cite{s2e-ttf1} of the TESLA Test Facility Free Electron Laser
(TTF FEL), Phase 1 \cite{ttf1}, have
shown a presence of a strong energy chirp (energy-time correlation) within a
short high-current leading peak in electron density distribution that has
driven Self-Amplified Spontaneous Emission (SASE) FEL process.
The energy chirp was accumulated due to the
longitudinal space charge after compression. According to the simulations (that
reproduced well the measured FEL properties), the energy chirp had a dramatical
impact on SASE FEL saturation length and output characteristics. A similar
effect takes place during the operation of VUV FEL at DESY in a "femtosecond
mode"  \cite{s2e-ttf2,stat-ttf2,ttf2-epj}. Such a mode of operation might also be possible in
future X-ray SASE FELs.

There also exists a concept of frequency-chirped SASE FELs (frequency
chirp of SASE FEL radiation is correlated with energy chirp in the electron
beam due to the FEL resonance condition) aiming at the shortening of radiation
pulse with the help of a monochromator \cite{schroeder}. Energy chirp can also
be used to tune the output frequency of an FEL with coherent prebunching as it
was demonstrated in the experiment at the DUV FEL facility \cite{duv}. Thus, a
theoretical understanding of the energy chirp effect on the FEL performance is
of crucial importance.

Analytical studies on this subject were performed in \cite{krinsky} in the framework of
one-dimensional approximation. The general form of a time-domain Green's
function as an inverse Laplace transform was derived in \cite{krinsky}. It was
then reduced to the explicit expression in the limit of small energy chirp
parameter up to the first order, resulting in phase correction (and ignoring
the gain correction). This explicit solution for the Green's function was used
to analyze statistical properties of a chirped SASE FEL in this limit. A second
order correction to the FEL gain was presented in \cite{schroeder} but this
result is incorrect.

In this paper we study the impact of energy chirp on SASE FEL
performance. We also find that FEL gain degradation can be perfectly compensated by undulator
tapering. We discuss an application of the compensation effect for generation of attosecond
pulses from X-ray FELs such as European XFEL \cite{xfel-tdr} and
Linac Coherent Light Source \cite{lcls}.

\section{Green's function}

 Let us consider a planar undulator with the magnetic field

\begin{displaymath}
H_z(z) =
H_{\mathrm{w}}\cos (2\pi z/\lambda _{\mathrm{w}}) \ ,
\end{displaymath}

\noindent where $\lambda _{\mathrm{w}}$ is undulator period, and
$H_{\mathrm{w}}$ is peak magnetic field.
Electric field of the amplified electromagnetic wave is presented in the
form:

\begin{displaymath}
E = \tilde{E} \exp [i \omega_0 (z/c - t )] + C.C. \ ,
\end{displaymath}

\noindent where $\omega_0$ is a reference frequency and $\tilde{E}$ is
slowly-varying amplitude \cite{book}. As it was shown in \cite{krinsky}, for a
SASE FEL, driven by an electron beam with linear energy chirp, $\tilde{E}$ can
be written as follows (we use notations from \cite{book}):

\begin{equation}
\tilde{E} = 2E_0 \sum_j e^{-i \hat{s}_j/\rho} e^{2i \hat{\alpha}\hat{s}_j (\hat{s}-\hat{z}/2-\hat{s}_j)} g(\hat{z}, \hat{s}-\hat{s}_j, \hat{\alpha})
\label{field}
\end{equation}

\noindent Here $\rho = \lambda _{\mathrm{w}} \Gamma/(4 \pi)$ is the efficiency parameter,
$\Gamma^3 = \pi j_0 K^2 A_{JJ}^2 /(I_{A} \lambda_{\mathrm{w}} \gamma_0^3)$,
$j_0$ is the beam current density, $I_{A} = mc^3/e \simeq 17$~kA, $\gamma_0$ is
relativistic factor,
$K = e \lambda_{\mathrm{w}} H_{\mathrm{w}} / (2\sqrt{2} \pi m c^2)$ is rms undulator parameter,
$A_{JJ} = J_0(Q)-J_1(Q)$ is the Bessel function factor, $Q=K^2/[2(1+K^2)]$,
$E_0 =  \rho \Gamma \gamma_0^2 m c^2 /(e K A_{JJ} \sqrt{2})$ is the
saturation field amplitude, $\hat{z} = \Gamma z$ is a normalized
position along the undulator, $\hat{s}=\rho\omega_0(z/\bar{v}_{z0}-t)$ is normalized
position along the electron bunch, $\bar{v}_{z0}$ is average longitudinal
velocity (defined for a reference particle). Let the energy linearly depend on
a particle position in the bunch (or arrival time). The energy chirp parameter

\begin{equation}
\hat{\alpha} = -\frac{d \gamma}{dt} \frac{1}{\gamma_0 \omega_0 \rho^2}
\label{eq:chirp-parameter}
\end{equation}

\noindent is defined such that, for positive sign of $\hat{\alpha}$, particles
in the head of the bunch have larger energy than those in the tail.
Relativistic factor $\gamma_0$ for a reference particle (placed at $\hat{s}=0$)
and reference frequency $\omega_0$ are connected by the FEL resonance
condition: $\omega_0 = 2ck_w \gamma_0^2/(1+K^2)$.
Note that the theory is applicable when $\rho \hat{\alpha} \ll 1$
\cite{krinsky}. It is also useful to define normalized detuning \cite{book}:
$\hat{C} = [k_w-\omega(1+K^2)/2c\gamma_0^2]/\Gamma$.

The Green's function $g$, entering Eq.~(\ref{field}), is given by the inverse
Laplace transform \cite{krinsky}:

\begin{equation}
g(\hat{z},\hat{s},\hat{\alpha}) =
2 \int \limits_{\gamma'-i \infty}^{\gamma'+i \infty} \frac{dp}{2\pi i p} \exp[f(p,\hat{z},\hat{s},\hat{\alpha})] \ ,
\label{eq:green-laplace}
\end{equation}

\noindent where

\begin{equation}
f(p,\hat{z},\hat{s},\hat{\alpha}) = p(\hat{z}-2\hat{s})+\frac{2i\hat{s}}{p(p+i\hat{\alpha}\hat{s})}
\label{eq:funct-f}
\end{equation}

\noindent We use a saddle point approximation to get an estimate of the
integral (\ref{eq:green-laplace}) for large values of $\hat{z}$ \cite{krinsky}.
The saddle point is determined from the condition $f'=0$ which leads to the 4th
power equation with three parameters:

\begin{equation}
p^4+2i\hat{\alpha}\hat{s}p^3-\hat{\alpha}^2 \hat{s}^2 p^2 - \frac{4i\hat{s}}{\hat{z}-2\hat{s}} \ p
+ \frac{2\hat{\alpha}\hat{s}^2}{\hat{z}-2\hat{s}} = 0
\label{dispersion}
\end{equation}

Once the saddle point, $p_0$, is found, the Green's function can be
approximated as follows:

\begin{equation}
g(\hat{z},\hat{s},\hat{\alpha}) =
\frac{2\exp[f(p_0,\hat{z},\hat{s},\hat{\alpha})]}{p_0 [2\pi f''(p_0,\hat{z},\hat{s},\hat{\alpha})]^{1/2}}
\label{eq:green-solution}
\end{equation}

Let us first consider the case when the energy chirp is a small perturbation,
$|\hat{\alpha}| \hat{z} \ll 1$, $\hat{z} \gg 1$. A second-order expansion of
the Green's function takes the following form

\begin{displaymath}
g(\hat{z},\hat{s},\hat{\alpha}) \simeq
\frac{e^{-i\pi/12}}{\sqrt{\pi \hat{z}}} \exp \left[ i^{1/3} \hat{z} + i^{2/3}\frac{\hat{\alpha}\hat{s}}{2} \left(1+i \frac{\hat{\alpha}\hat{z}^2}{36}  \right)
\right.
\end{displaymath}
\begin{equation}
\left. -9 i^{1/3} \left( 1- \frac{\hat{\alpha}^2 \hat{z}^2}{216 i^{2/3}} \right) \frac{(\hat{s}-\hat{z}/6)^2}{\hat{z}}
-\frac{i}{2} \hat{\alpha}\hat{s} (\hat{z}-2\hat{s})  \right]
\label{green-smallchirp}
\end{equation}

\noindent The leading correction term is the last term in the argument of the
exponential function. This term was found in \cite{krinsky} (note difference in
definition of normalized parameters). Setting $\hat{\alpha}=0$, one gets from
(\ref{green-smallchirp}) the well-known Green's function for unchirped beam
\cite{krinsky-stat}.

Now let us consider the case $\hat{\alpha} > 0$ and  $1 \ll \hat{\alpha} \ll
\hat{z}$. The Green's function for $\hat{s} \gg \hat{\alpha}^{-1}$ is
approximated by:

\begin{equation}
g(\hat{z},\hat{s},\hat{\alpha}) \simeq
\left( \frac{\hat{\alpha}}{2\pi^2 \hat{z}} \right)^{1/4}
\exp \left( 2 \sqrt{\frac{2\hat{z}}{\hat{\alpha}}}
- 2 \sqrt{\frac{2}{\hat{\alpha} \hat{z}}} \ \hat{s} \right)
\label{green-large-positive}
\end{equation}

\noindent More thorough analysis for small values of $\hat{s}$ shows that the
Green's function has a maximum at $\hat{s}_{\mathrm{m}}=2^{1/3}\hat{\alpha}^{-1}$,
i.e. the position of maximum is independent of $\hat{z}$ while the width of the
radiation wavepacket is proportional to $\sqrt{\hat{\alpha} \hat{z}}$. The mean
frequency of the radiation wavepacket corresponds to a resonant frequency at
$\hat{s}=0$. Note also that the beam density excitation is concentrated near
$\hat{s}=0$ within much shorter range, of the order of
$\hat{\alpha}^{-7/4}\hat{z}^{-1/4}$.

In the case of $\hat{\alpha} < 0$ and  $1 \ll |\hat{\alpha}| \ll \hat{z}$ the
Green's function is given by:

\begin{displaymath}
g(\hat{z},\hat{s},\hat{\alpha}) \simeq
\frac{2^{1/4}e^{-i\pi/2}}{\pi^{1/2}|\hat{\alpha}|^{5/4} \hat{z}^{3/4} \hat{s}}
\exp \left( 2 \sqrt{\frac{2\hat{z}}{|\hat{\alpha}|}} \right.
\end{displaymath}
\begin{equation}
\left. + i|\hat{\alpha}|\hat{z}\hat{s} +\frac{2 i}{|\hat{\alpha}|^2\hat{s}} - \frac{2\sqrt{2}}{|\hat{\alpha}|^{7/2} \hat{z}^{1/2} \hat{s}^2}
- 2 \sqrt{\frac{2}{|\hat{\alpha}| \hat{z}}} \ \hat{s}
\right)
\label{green-large-negative}
\end{equation}

\noindent The width of the radiation wavepacket (and of the beam density excitation
as well) is of the order of $|\hat{\alpha}|^{-7/4}\hat{z}^{-1/4}$. The maximum
of the wavepacket is positioned at $\hat{s}_{\mathrm{m}} =
2^{5/4}|\hat{\alpha}|^{-7/4}\hat{z}^{-1/4}$, i.e. the wavepacket is shrinking
and back-propagating (with respect to the electron beam) with increasing
$\hat{z}$. The mean frequency of the wavepacket is blue-shifted with respect to
resonant frequency at $\hat{s} = 0$. In normalized form this shift is $\Delta
\hat{C} = - |\hat{\alpha}| \hat{z}/2$.

\section{Linear regime of SASE FEL}

The normalized radiation power (normalized efficiency), $<\hat{\eta}> =
P_{\mathrm{SASE}}/\rho P_{\mathrm{beam}}$, can be expressed as follows
\cite{book}:

\begin{equation}
<\hat{\eta}> = \frac{<|\tilde{E}|^2>}{4 E_0^2} \ ,
\label{eta}
\end{equation}

\noindent where $<...>$ means ensemble average. One can easily get from
(\ref{field}):

 \begin{equation}
<\hat{\eta} (\hat{z},\hat{\alpha})> = \frac{1}{N_{\mathrm{c}}} \int_0^{\infty} d \hat{s} |g(\hat{z},\hat{s},\hat{\alpha})|^2 \ .
\label{eta-green}
\end{equation}

\noindent Here $N_{\mathrm{c}}=N_{\lambda}/(2\pi \rho)$ is a number of
cooperating electrons (populating $\Delta \hat{s} = 1$), $N_{\lambda}$ is a
number of electrons per wavelength. The local power growth rate \cite{huang}
can be computed as follows:

\begin{equation} G(\hat{z},\hat{\alpha}) = \frac{d}{d\hat{z}} \ln  <\hat{\eta}
(\hat{z},\hat{\alpha})> \ . \label{diff-growth} \end{equation}

Applying Eqs. (\ref{eta-green}), (\ref{diff-growth}) to the asymptotical cases,
considered in the previous Section, we get the following results. For the case
$|\hat{\alpha}| \hat{z} \ll 1$, $\hat{z} \gg 1$ the FEL power is given by

\begin{equation}
<\hat{\eta}> \simeq
\frac{\exp \left\{ \sqrt{3} \hat{z} \left[ 1- \left( \hat{\alpha}\hat{z}/12 \right)^2/3 \right] + \hat{\alpha}\hat{z}/12 \right\}}
{3^{5/4} \sqrt{\pi \hat{z}} N_{\mathrm{c}}}
\label{eta-small}
\end{equation}

\noindent and the local power growth rate is

\begin{equation}
G(\hat{z},\hat{\alpha}) \simeq \sqrt{3} \left[1- \left( \frac{\hat{\alpha}\hat{z}}{12} \right)^2 \right] - \frac{1}{2\hat{z}} + \frac{\hat{\alpha}}{12} \ .
\label{diff-growth-smallchirp}
\end{equation}

\noindent It reaches maximum $G_{\mathrm{m}} = \sqrt{3} \left[ 1- \left(
|\hat{\alpha}|/16 \right)^{2/3}  \right] + \hat{\alpha}/12$ at the position
$\hat{z}_{\mathrm{m}}=3^{1/2}2^{2/3}/|\hat{\alpha}|^{2/3}$. Numerical simulations show
that Eqs.~(\ref{eta-small}) and (\ref{diff-growth-smallchirp}) are pretty
accurate up to the values $|\hat{\alpha}| \hat{z}$ of the order of unity although
the condition $|\hat{\alpha}| \hat{z} \ll 1$ was used to derive them.

For the case $\hat{\alpha} > 0$ and  $1 \ll \hat{\alpha} \ll \hat{z}$ we get
rather simple expressions:

\begin{equation}
<\hat{\eta} (\hat{z},\hat{\alpha}) > \simeq
\frac{\hat{\alpha}}{8\pi N_{\mathrm{c}}}
\exp \left( 4 \sqrt{ \frac{2\hat{z}}{\hat{\alpha}} } \right) ,
\label{eta-large-pos}
\end{equation}

\begin{equation}
G(\hat{z},\hat{\alpha}) \simeq 2 \sqrt{\frac{2}{\hat{\alpha}\hat{z}}}  \ .
\label{diff-growth-large-positive}
\end{equation}

For large negative values of $\hat{\alpha}$ we obtain:

\begin{equation}
<\hat{\eta}> \simeq
\frac{1}{2^{7/4} \pi^{1/2} |\hat{\alpha}|^{3/4} \hat{z}^{5/4} N_{\mathrm{c}}}
\exp \left( 4 \sqrt{ \frac{2\hat{z}}{|\hat{\alpha}|} } \right) ,
\label{eta-large-neg}
\end{equation}

\begin{equation}
G(\hat{z},\hat{\alpha}) \simeq 2 \sqrt{\frac{2}{|\hat{\alpha}|\hat{z}}} - \frac{5}{4\hat{z}} \ .
\label{diff-growth-large-negative}
\end{equation}

\section{Nonlinear regime}

We studied nonlinear regime of a chirped SASE FEL operation with 1-D version of
the code FAST \cite{book,fast}. Analytical results, presented above,
were used as a primary standard for testing the code in linear regime. Green's function
was modelled by exciting density modulation on a short scale, $\Delta \hat{s}
\ll 1$. SASE FEL initial conditions were simulated in a standard way
\cite{book}. The results of numerical simulations in all cases were in a good
agreement with analytical results presented in two previous Sections.

\begin{figure}[tb]

\includegraphics[width=0.7\textwidth]{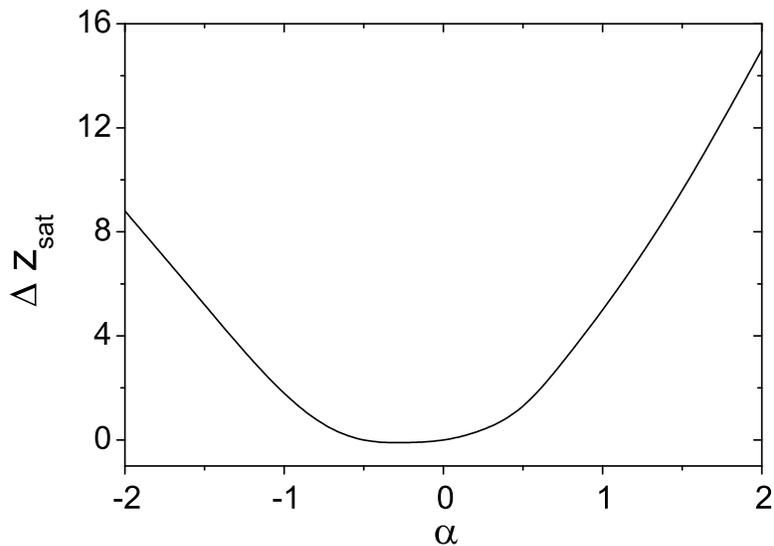}

\caption{
Increase of saturation length $\Delta \hat{z}_{\mathrm{sat}} =
\hat{z}_{\mathrm{sat}}(\hat{\alpha})- \hat{z}_{\mathrm{sat}}(0)$ versus
parameter $\hat{\alpha}$. Here $\hat{z}_{\mathrm{sat}}(0) = 13$.
}
\label{fig:f1}
\end{figure}

\begin{figure}[tb]

\includegraphics[width=0.7\textwidth]{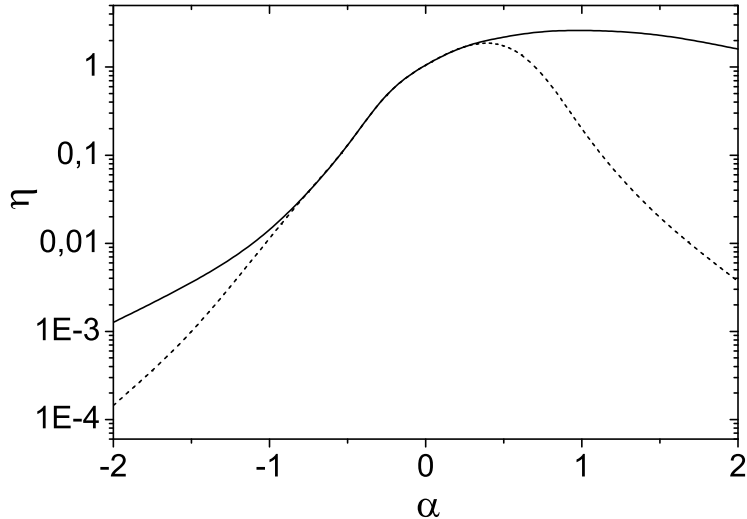}

\caption{
Normalized output power versus parameter $\hat{\alpha}$. Solid: $\hat{z}=
\hat{z}_{\mathrm{sat}} (\hat{\alpha})$ (see Fig.~\ref{fig:f1}); dash: $\hat{z}=
\hat{z}_{\mathrm{sat}} (0) =13$.
}
\label{fig:f2}
\end{figure}

The main results of
the simulations of the nonlinear regime are presented in Figs.~\ref{fig:f1} and
\ref{fig:f2}. Saturation length and power are functions of two parameters,
$\hat{\alpha}$ and $N_{\mathrm{c}}$. For our simulations we have chosen
$N_{\mathrm{c}}=3\times 10^7$ - a typical value for VUV SASE FELs. Note,
however, that the results, presented in Figs.~\ref{fig:f1} and \ref{fig:f2},
very weakly depend on $N_{\mathrm{c}}$. Fig.~\ref{fig:f1} shows increase of
saturation length with respect to unchirped beam case. In Fig.~\ref{fig:f2} the
output power is plotted versus chirp parameter for two cases: when undulator
length is equal to a saturation length for a given $\hat{\alpha}$ and when it
is equal to the saturation length for the unchirped beam case. One can see
sharp reduction of power for negative $\hat{\alpha}$ while a mild positive
chirp ($\hat{\alpha} < 0.5$) is beneficial for SASE.

\section{Energy chirp and undulator tapering}

\begin{figure}[tb]

\includegraphics[width=0.7\textwidth]{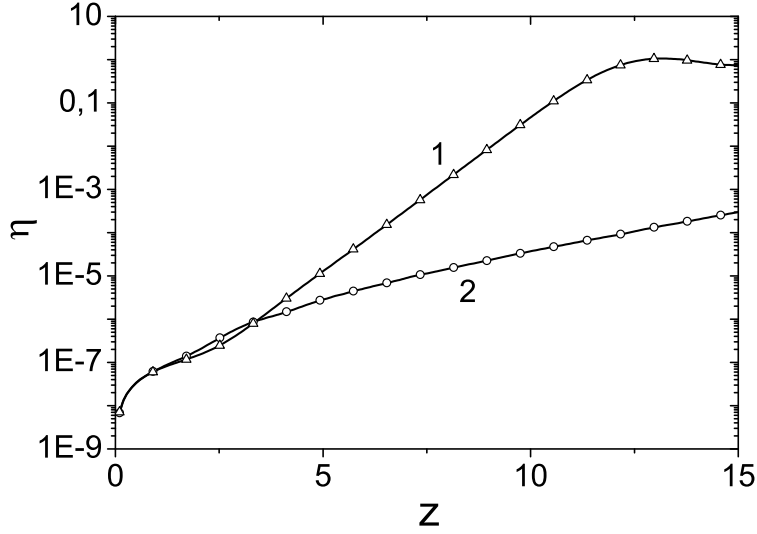}

\caption{
Normalized power versus undulator length. Solid line 1: $\hat{\alpha}=0$,
$\hat{b}_1=0$; triangles: $\hat{\alpha}=4$, $\hat{b}_1=-2$; solid line 2:
$\hat{\alpha}=4$, $\hat{b}_1=0$; circles: $\hat{\alpha}=0$, $\hat{b}_1=2$.
}
\label{fig:f3}
\end{figure}

Let us consider now the case when there is no energy chirp ($\hat{\alpha}=0$)
and the detuning parameter changes linearly along the undulator \cite{book}:
$\hat{C}(\hat{z})= \hat{b}_1 \hat{z}$. This change can be due to variation of
undulator parameters ($K(\hat{z})$ and/or $k_{\mathrm{w}}(\hat{z}$)), or due to
an energy change $\gamma_0(\hat{z})$. We have found from numerical simulations
that in such case the effect on FEL gain is exactly the same as in the case of
energy chirp and no taper if $\hat{\alpha}=2\hat{b}_1$ for any value of
$\hat{\alpha}$ (Figure~\ref{fig:f3} shows an example). Therefore, all the
results of two previous Sections can be also used for the case of linear
variation of energy or undulator parameters with the substitution $\hat{\alpha}
\to 2\hat{b}_1$. The amplitudes of Green's functions are also the same while
the phases are obviously different. In case of $\hat{b}_1=0$, $\hat{\alpha} \ne
0$ there is a frequency chirp along the bunch while in the case $\hat{b}_1=0$,
$\hat{\alpha} \ne 0$ the frequency is changing along the undulator.

An effect of undulator tapering (or energy change along the undulator) on FEL
gain was studied in \cite{huang} in the limit $\hat{b}_1 \ll 1$. Comparing our
Eq.~(\ref{eta-small}) (with the substitution $\hat{\alpha} \to 2\hat{b}_1$) and
Eq.~(45) of Ref.~\cite{huang}, we can see that quadratic correction term in the
argument of the exponential function is the same but the linear term is two
times larger in \cite{huang}. The reason for discrepancy is that the frequency
dependence of the pre-exponential factor in Eq.~(42) of Ref.~\cite{huang} is
neglected.

A symmetry between two considered effects (energy chirp and undulator tapering)
can be understood as follows. If we look at the radiation field acting on some
test electron from an electron behind it, this field was emitted at a retarded
time. In the first case a back electron has a detuning due to an energy
offset, in the second case it has the same detuning because undulator
parameters were different at a retarded time. The question arises: can these
two effects compensate each other? We give a positive answer based on numerical
simulations (see Fig.~\ref{fig:f3} as an example): by setting $\hat{b}_1 = -
\hat{\alpha}/2$ we get rid of gain degradation, and FEL power at any point
along the undulator is the same as in the case of unchirped beam and untapered
undulator. This holds for any value of $\hat{\alpha}$. For instance, if one
linearly changes magnetic field $H_w$ of the undulator, the compensation
condition can be written as follows (nominal values of parameters are marked
with subscript '0'):

\begin{equation}
\frac{1}{H_{\mathrm{w} 0}} \frac{d H_{\mathrm{w}}}{dz} = -\frac{1}{2} \frac{(1+K_0^2)^2}{K_0^2} \frac{1}{\gamma_0^3} \frac{d \gamma}{c dt}
\label{compensation}
\end{equation}

\noindent Of course, in such a case we get frequency chirped SASE pulse. Since
compensation of gain degradation is possible also for large values of
$\hat{\alpha}$ (there is no theoretical limit on the value of chirp parameter,
except for above mentioned condition $\rho \hat{\alpha} \ll 1$), one can, in
principle, organize a regime when a frequency chirp within an intensity spike
is much larger than the natural FEL bandwidth (given by $\rho \omega_0$).

\section{Generation of attosecond pulses}

Up to now several schemes for generation of attosecond pulses from X-ray SASE
FELs have been proposed \cite{atto-a,atto-b,atto-c,atto-d,atto-e,atto-f}.
Here we mention the schemes considered in
\cite{atto-c,atto-d} making use of energy modulation of a short slice in the
electron bunch by a high-power few-cycle optical pulse in a two-period
undulator. Due to energy modulation the frequency of SASE radiation in X-ray
undulator is correlated to the
longitudinal position within the few-cycle-driven slice of the electron beam.
The largest frequency offset corresponds to a single-spike pulse in time domain
(about 300 attoseconds). The selection of single-spike pulses is
achieved by using a crystal monochromator after the X-ray undulator \cite{atto-c},
or with the help of the other undulator tuned to the offset frequency \cite{atto-d}.

\begin{figure}[tb]

\includegraphics[width=0.9\textwidth]{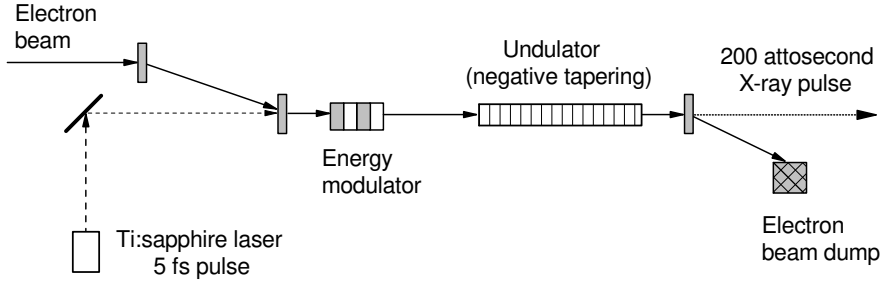}

\caption{
Schematic diagram of attosecond X-ray source.
Energy modulator performs slice energy modulation of
the electron bunch (see Fig.~\ref{fig:emod40}). Undulator
tapering leads to complete suppression of the amplification
process in the most fraction of the electron bunch, and
output X-ray pulse has 200 attosecond pulse duration.
}
\label{fig:atto-chirp}
\end{figure}

In this paper we propose a new scheme (see Fig.~\ref{fig:atto-chirp}) that makes use of
the compensation effect, described in the previous Section. Indeed, there is
a strong energy chirp around zero-crossing of energy modulation. If one uses
appropriate undulator taper then only a
short slice around zero-crossing produces powerful FEL pulse. The main part of
the bunch is unmodulated and suffers from strong negative
undulator tapering (see Fig.~\ref{fig:f2}). One should also note that for large negative
taper the SASE FEL gain is very sensitive to longitudinal velocity spread.
Therefore, a high-contrast attosecond pulse is directly
produced in the undulator.

\begin{figure}[tb]

\includegraphics[width=0.7\textwidth]{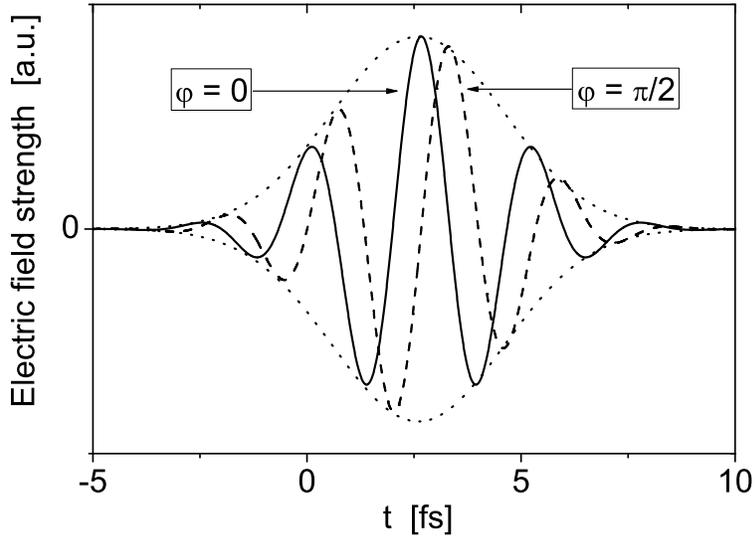}

\caption{
Possible evolutions of the electric field in the 5-fs pulse.
carried at a wavelength 800 nm for two different pulse phases ($\phi =
0, \pi/2$)
}
\label{fig:lf}
\end{figure}

\begin{figure}[tb]

\includegraphics[width=0.7\textwidth]{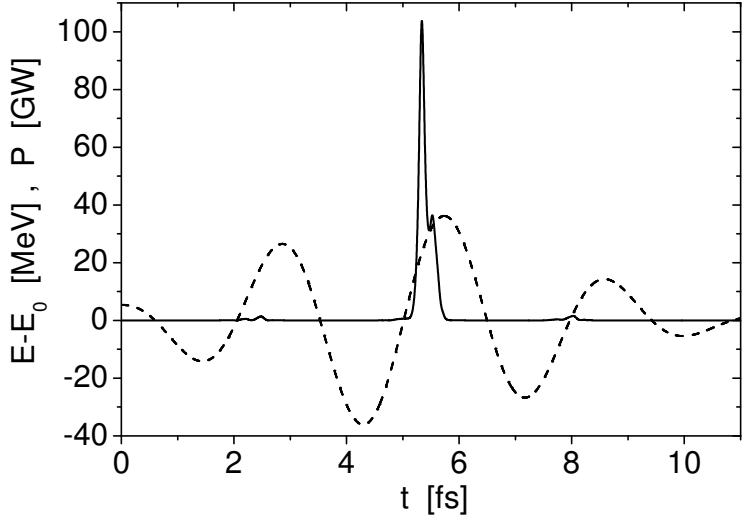}

\caption{
Energy modulation of the electron beam at the exit of the
modulator undulator (dotted line) and a  profile of the
radiation pulse at the undulator length 100~m
}
\label{fig:emod40}
\end{figure}

Operation of attosecond SASE FEL is illustrated for the
parameters close to those of the European XFEL operating at the
wavelength 0.15 nm \cite{xfel-tdr}.
The parameters of the electron beam
are: energy 15~GeV, charge 1~nC, rms pulse length 25~$\mu $m, rms
normalized emittance 1.4~mm-mrad, rms energy spread 1~MeV. Undulator
period is 3.65~cm.

The parameters of the seed laser are: wavelength 800 nm, energy in the
laser pulse 3 mJ, and FWHM pulse duration 5 fs (see Fig.
\ref{fig:lf}). The laser beam is focused onto the electron beam
in a short undulator resonant at the optical wavelength of 800~nm.
Optimal conditions of the focusing correspond to the positioning of the
laser beam waist in the center of the modulator undulator. It is
assumed that the phase of laser field corresponds to "sine" mode
(dashed line with $ \varphi = \pi /2$, see Fig.~\ref{fig:lf}).
Parameters of the modulator undulator are: period length 50~cm, peak
field 1.6~T, number of periods 2. The interaction with the laser light
in the undulator produces a time-dependent electron energy modulation
as it is shown in Fig.~\ref{fig:emod40}. This modulation
corresponds to the energy chirp parameter $\hat{\alpha} \simeq 2$
at zero crossing ($t = 5$ fs in Fig.~\ref{fig:emod40}).

Optimization of the attosecond SASE FEL has been performed
with the three-dimensional, time dependent code FAST \cite{fast} taking
into account all physical effects influencing the SASE FEL operation
(diffraction effects, energy spread, emittance, slippage effect, etc.).
Three-dimensional simulations confirmed the predictions of
the one-dimensional model: the energy chirp and the undulator tapering compensate
each other, there is strong suppression of the amplification in the case of
uncompensated negative taper.

Undulator tapering is performed by changing the gap of undulator modules \cite{xfel-tdr}
such that magnetic field increases linearly along the undulator length ($\hat{b}_1 < 0$).
We performed the scan of tapering depth $\hat{b}_1$ in order to maximize the power in the
main peak on one hand,
and to minimize contribution of the background, on the other hand.
We ended up with the value of taper which is
about 20 \% smaller than that required for a perfect compensation of chirp at $t = 5$ fs.
Note that the chirp is not linear in the region of interest. In addition, a mild net positive chirp
is beneficial for SASE, as it was discussed above (see Fig.~\ref{fig:f2}).

\begin{figure}[tb]

\includegraphics[width=0.7\textwidth]{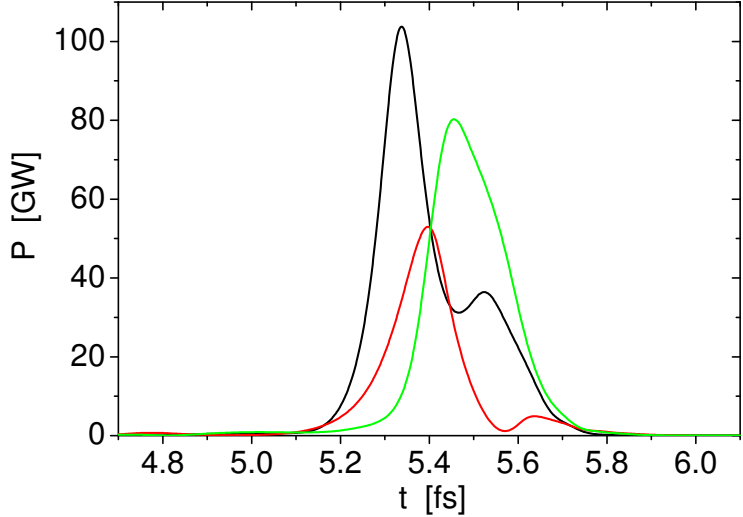}

\caption{
Temporal structure of the
radiation pulse (three different shots)
at the undulator length 100~m.
}
\label{fig:ps0192050}
\end{figure}

A typical radiation pulse at the undulator length 100 m is shown
in Fig.~\ref{fig:emod40}. One can see a high-power spike in the region where the energy chirp
is well compensated by the taper and two weak side peaks at $t \simeq 2$ fs and $t \simeq 8$ fs
where the net effect is negative taper. In the rest of the bunch a large negative taper
together with velocity spread and 3-D effects completely suppresses amplification.
In Fig.~\ref{fig:ps0192050} we present three different shots illustrating
the properties of the main peak. Typical pulse duration is about 200 attoseconds (FWHM) and
peak power ranges from several tens up to hundred GW.
To estimate the contrast (which we define as the ratio of energy in the main peak to the
total radiated energy at the experiment) we assume that an angular collimation is
used in order to reduce
spontaneous emission background. A collimator with half-angle 3 $\mu$rad allows
the entire intensity in the main peak to be transmitted. The contrast is influenced
by SASE intensity in two side peaks and by spontaneous emission in the first harmonic
from the rest of the bunch. For the charge of 1 nC, as in our numerical example,
the contrast is about 95 \%. Higher harmonics
of undulator radiation (if they disturb an experiment) can be cut, for instance, by a multilayer
monochromator with a bandwidth of the order of 1 \%.

\section{Beyond "fundamental limit"}

It is generally accepted that the shortest pulse, that can be obtained from a SASE FEL,
is given by a duration of intensity spike in time domain, i.e. it is defined by inverse FEL
bandwidth $(\rho \omega_0)^{-1}$.
However, the fact that a SASE FEL can operate with a strong chirp parameter (in
combination with undulator tapering) without gain degradation, opens up a
possibility of a conceptual breakthrough: one can get from SASE FEL a radiation
pulse which is much shorter than the inverse FEL bandwidth. Indeed, in the case of
$\hat{\alpha} \gg 1$, the frequency chirp inside an intensity spike
is much larger than FEL bandwidth. Thus, one can use a monochromator to reduce pulse duration.
By an appropriate choice
of the monochromator bandwidth one can select an X-ray pulse that
is shorter by a factor of $\sqrt{2\hat{\alpha}}$ than the inverse FEL bandwidth.
The only theoretical
limit in this case is given by the condition $\rho \hat{\alpha} \ll 1$. Note that for hard X-ray
FELs the parameter $\rho$ is in the range $10^{-4}-10^{-3}$.

To illustrate a possible technical realization of this idea, we can suppose
that the energy modulation by a few-cycle optical pulse
is increased by a factor 3 so that $\hat{\alpha} \simeq 6$.
In combination with undulator tapering and a monochromator, this would allow to
obtain sub-100-GW coherent X-ray pulses with a duration below 100 attoseconds
and a contrast about 80-90 \%.

\bibliography{chirp-prst}

\end{document}